\documentclass{jetplm}
\twocolumn

\usepackage{xcolor}
\usepackage{amsthm}
\usepackage{algorithm}
\usepackage[noend]{algpseudocode}
\usepackage[utf8]{inputenc}
\usepackage[T2A]{fontenc}
\usepackage[russian,english]{babel}

\DeclareMathOperator{\argmin}{argmin}

\newtheoremstyle{exampstyle}
  {0pt} 
  {0pt} 
  {} 
  {} 
  {\bfseries} 
  {.} 
  {.5em} 
  {} 
\theoremstyle{exampstyle}
\newtheorem*{theorem}{Theorem}

\lat

\title{Quantum $\mathcal{R}$-matrices as universal qubit gates}

\rtitle{ Quantum $\mathcal{R}$-matrices as universal qubit gates}

\sodtitle{ Quantum $\mathcal{R}$-matrices as universal qubit gates}

\author{N. Kolganov$^{ab}$\thanks{e-mail: nikita.kolganov@phystech.edu},
An. Morozov$^{abc}$\thanks{e-mail: andrey.morozov@itep.ru}}

\rauthor{N. Kolganov, An. Morozov}

\sodauthor{N. Kolganov, An. Morozov}

\address{$^a$ {\small {\it MIPT, Dolgoprudny, 141701, Russia}}\\
$^b$ {\small {\it ITEP, Moscow 117218, Russia}}\\
$^c$ {\small {\it Institute for Information Transmission Problems, Moscow 127994, Russia}}
}

\dates{ }{ }

\abstract{We study the Chern-Simons approach to the topological quantum computing. We use quantum $\mathcal{R}$-matrices as universal quantum gates and study the approximations of some one-qubit operations. We make some modifications to the known Solovay-Kitaev algorithm suitable for our particular problem.}

\PACS{ }

\begin{document}

\maketitle

\section{Introduction}

Quantum computers is quite a hot topic nowadays. They are a very promising device and method of solving lots of problems. The main problem of the quantum computers from the practical point of view is the high probability of errors. Since it has quantum nature the states are not stable and can dissolve or change. This limits the time the quantum computer can work and the size of the programs it can run. To deal with these problems the quantum correction algorithms are usually used. These algorithms imply that instead of physical qubits the logical qubits, consisting of several physical ones, are considered. The physical qubits inside of the logical qubit are regularly entangled with each other thus providing an error corrections. However, the quntum computer with such error corrections require many more physical qubits which is also a big problem at the current stage.

Another approach is to try to use qubit models where the states are more stable. One of the approaches is to make the states topological, as those are usually much harder to change. This leads to the idea of the topological quantum computer. Many of the models of this quantum computer behave under the laws of the topological 3d Chern-Simons theory. There are different models where the Chern-Simons is an effective theory which in future could provide us with the topological quantum computer\footnote{Due to vastness of literature on the quantum computer in general and topological quantum computer in particular we give here only a link to a review on topological quantum computing \cite{tqc,tqc1}.}.

The main observables, studied in the Chern-Simons theory are Wilson-loop averages, and this loops are usually thought to be related to quantum programs or algorithms. As we know \cite{Witt} these Wilson-loop averages are equal to the knot invariants. According to the Reshetikhin-Turaev approach \cite{RT}-\cite{RTfin}, these knot invariants can be constructed from the $\mathcal{R}$-matrices. In this sense these matrices provide an elementary building blocks from which the whole knot is constructed. In quantum information theory such building blocks are called universal quantum gates. In \cite{tqcKauf,tqcus} it was suggested that quantum $\mathcal{R}$-matrices can indeed be used as universal quantum gates.

In the present papers we continue these studies with the goal of studying the properties of the $\mathcal{R}$-matrices as quantum gates and how different other gates can be constructed from them using Solovay-Kitaev algorithm \cite{kitaev, dawson}. We construct an approximation for one-qubit Hadamard and $\pi/8$ gates from fundamental $\mathcal{R}$ and Racah matrices. The generalization to larger matrices and higher representations is a work in progress.

\section{3d Chern-Simons theory}

The simplest topological theory is Chern-Simons theory. It is a 3-dimensional topological gauge theory with action

\begin{equation}\label{CSaction}
\begin{array}{l}
  S=\frac{k}{4\pi} \int \text{Tr}\left( \mathcal{A}\wedge d\mathcal{A}+\frac{2}{3}\mathcal{A}\wedge\mathcal{A}\wedge\mathcal{A}\right)=
  \\
  = \frac{k}{4\pi} \int d^3x \epsilon^{ijk}\left(\delta_{ab} \mathcal{A}^a_i\partial_j\mathcal{A}^b_k+\frac{2}{3}f_{abc}\mathcal{A}^a_i\mathcal{A}^b_j\mathcal{A}^c_k\right).
\end{array}
\end{equation}

The most interesting observables of such theory are Wilson-loop averages

\begin{equation}\label{WL}
  \left<W^{\mathcal{K}}_T\right>_{CS}=\left< \text{Tr}_{T} \text{Pexp} \oint_{\mathcal{K}} \mathcal{A} \, dx\right>_{CS}
\end{equation}

We know that these Wilson-loop averages for the $SU(N)$ gauge group are equal to HOMFLY-PT polynomials \cite{Witt}. HOMFLY-PT polynomials depend on two variables $A$ and $q$, which are connected to the parameters of the theory:

\begin{equation}
q=\exp{\frac{2 \pi i}{k+N}},\ \ \ \ \ A=q^N.
\end{equation}

Gauge invariance of the Chern-Simons theory allows us to choose the gauge. The temporal gauge $\mathcal{A}_0=0$ leads to a representation of the Wilson-loops averages as a product of quantum $\mathcal{R}$-matrices \cite{RT}-\cite{RTfin}.

\section{HOMFLY-PT polynomials and quantum $\mathcal{R}$-matrices}

Quantum $\mathcal{R}$-matrix is a solution to the Yang-Baxter equation:

\begin{equation}
  \mathcal{R}_1\mathcal{R}_2\mathcal{R}_1=\mathcal{R}_2\mathcal{R}_1\mathcal{R}_2.
\end{equation}
This equation has a graphic representation as a third Reidemeister move:

\begin{center}
\begin{picture}(150,50)(70,-50  )
\put(-170,-60){
\put(250,10){\line(1,1){40}}
\put(250,50){\line(1,-1){17}}
\put(273,27){\line(1,-1){17}}

\put(257,53){\line(0,-1){7}}
\put(257,40){\line(0,-1){20}}
\put(257,14){\line(0,-1){7}}

\put(257,30){\vector(0,1){2}}
\put(280,40){\vector(1,1){2}}
\put(280,20){\vector(1,-1){2}}

\put(300,28){\mbox{$=$}}

\put(320,10){\line(1,1){40}}
\put(320,50){\line(1,-1){17}}
\put(343,27){\line(1,-1){17}}

\put(353,53){\line(0,-1){7}}
\put(353,40){\line(0,-1){20}}
\put(353,14){\line(0,-1){7}}

\put(353,30){\vector(0,1){2}}
\put(330,40){\vector(1,-1){2}}
\put(330,20){\vector(1,1){2}}}
\label{rand}
\end{picture}
\end{center}

For $\mathcal{R}$-matrix there is a known general formula \cite{RT4}:
\begin{equation}
\mathcal{R} = \mathcal{P} q^{\sum_{i,j} a^{-1}_{i,j} h_i \otimes h_j} \overrightarrow{\prod_{\beta \in \Phi^{+}}} {\rm exp}_{q}     \left ( (q - q^{-1}) E_{\beta} \otimes F_{\beta}   \right),
\end{equation}
where $E_{\beta}$, $F_{\beta}$ and $h_{\beta}$ are the generators of the quantum group.

One of the crucial properties of this universal $\mathcal{R}$-matrix is that it by definition commutes with rising $E$ and lowering $F$ operators. This means that it acts in the same way on all elements of the representation of the corresponding quantum group. Thus one can move from the universal quantum $\mathcal{R}$-matrix to the $\mathcal{R}$-matrix in the space of intertwining operators. This type of $\mathcal{R}$-matrix describes how it acts on all irreducible representations in the tensor product of a pair of representations. The eigenvalues of such a $\mathcal{R}$-matrices are described in a very simple way for any representation $Q$ from the product of two representations $R$:

\begin{equation}
\lambda_Q=\epsilon_Q q^{\varkappa_Q-4\varkappa_R}A^{-|R|},
\end{equation}
where $|R|$ is the size of the Young diagram $R$ and $\varkappa_R$ is
\begin{equation}
\varkappa_R=\sum_{\{i,j\}\in R} (i-j).
\end{equation}

The physically meaningful Chern-Simons theories has integer $k$ and of course integer $N$. This means that both $q$ and $A$ are roots of unity. Thus the $\mathcal{R}$-matrix in the space of intertwining operators is unitary.

\section{Two-bridge knots}

The important subclass of knots are two-bridge knots, also known as 4-plat and rational knots. These are knots which have two bridges -- if one draws this knot on a plane it requires only two arcs to be above the plain. Another definition of these knots is to take a 4-strand braid and then close it on each end, like this:

\begin{center}
\includegraphics[scale=0.2]{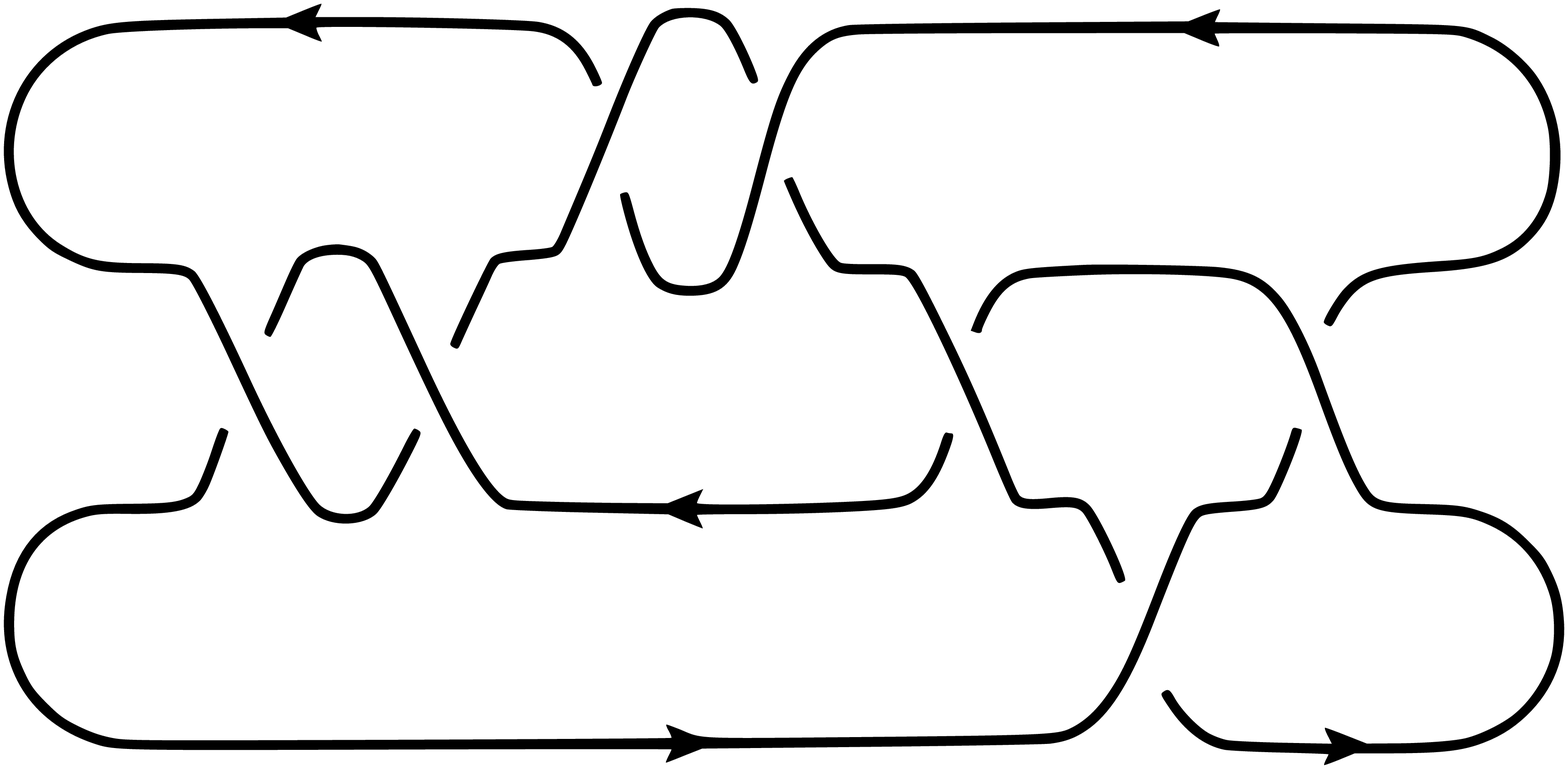}
\end{center}

From the point of view of Chern-Simons theory and representation theory strands going in one direction correspond to representation $R$ and strands going in the opposite direction correspond to the conjugate representation $\bar{R}$. Thus closures on both right and left ends correspond to the projection from the product $R\otimes \bar{R}$ onto the trivial representation $\emptyset$ which always exist in this product by the definition of the conjugate representation. Since $\mathcal{R}$-matrices do not change the overall representation, it means that overall representation in the product $R\otimes R\otimes \bar{R}\otimes \bar{R}$ should be chosen as $\emptyset$. This also leads to the close relationship between 3d Chern-Simons theory and 2d conformal field theory \cite{Witt,Rama}.

The HOMFLY-PT polynomials of such knots are equal to the element of the product of $\mathcal{R}$-matrices divided by the quantum dimension of the representation $R$. There are four different matrices which combinations give the answers for two-bridge knots. There are two diagonal $\mathcal{R}$-matrices, $T$ and $\bar{T}$, corresponding to two different types of crossings. Then there are two Racah matrices, $S$ and $\bar{S}$, which change the bases, so that we can diagonalize $\mathcal{R}$-matrices in different positions.

\section{Universal quantum gates and Solovay-Kitaev algorithm}

If we want to use quantum computer in an effective way we want to be able to run any program on such a computer. But to be able to work with such a computer there should be a finite number of elementary operations. If one can construct any program with any accuracy using some set of such elementary operations then this set is called a set of \textit{universal quantum gates}. There are different sets of universal quantum gates for different number of spins and different models of quantum computers.

There is a Solovay-Kitaev theorem \cite{kitaev}, which says that if there is a set of universal quantum gates then any program represented by unitary matrix, can be approximated in a logarithmic time and with logarithmic number of operators. It is formulated as follows.

\begin{theorem}[Solovay-Kitaev]
Let $\mathcal G$ be a set of universal gates, and let a desired accuracy $\epsilon > 0$ be fixed. There is a constant $c$ such that for any special unitary $U$ there is a finite sequence $S$ of gates from $\mathcal G$ of length $O(\log^c(1/\epsilon))$ and such that $||U - S|| < \epsilon$.
\end{theorem}

Here we assume that $g^\dagger$ is also in $\mathcal G$ for any $g \in \mathcal G$, and $||\cdot||$ stands for operator norm (largest singular value).
This theorem can be proved, e.g., by providing a specific algorithm constructing, which we refer as Solovay-Kitaev algorithm, of such an approximation \cite{dawson}. The pseudocode of this algorithm reads
\begin{algorithm}[H]
\caption{Solovay-Kitaev algorithm}
\begin{algorithmic}
    \Function{Solovay-Kitaev}{gate $U$, depth $n$}
    \If {$n = 0$}
        \State \Return \Call{Basic-Approximation}{$U$}
    \Else
        \State $U_{n-1} \gets \text{\Call{Solovay-Kitaev}{$U$, $n-1$}}$
        \State $V, W \gets \text{\Call{GC-Decompose}{$U U_{n-1}^\dagger$}}$
        \State $V_{n-1} \gets \text{\Call{Solovay-Kitaev}{$V$, $n-1$}}$
        \State $W_{n-1} \gets \text{\Call{Solovay-Kitaev}{$W$, $n-1$}}$
        \State \Return $V_{n-1} W_{n-1} V_{n-1}^\dagger W_{n-1}^\dagger U_{n-1}$
    \EndIf
    \EndFunction
\end{algorithmic}
\end{algorithm}
The function \textsc{Basic-Approximation}($U$) returns a sequence $U_0$ from the set of universal gates $\mathcal G$ that provide basic (zeroth order) approximation of $U$ with accuracy $\epsilon_0$, which should be sufficient for the convergence of the algorithm. This is usually achieved by brute force search of the best approximation among all sequences of gates $\mathcal G$ up to some fixed length $l_0$.  The function \textsc{GC-Decompose}($\Delta$) returns the so-called balanced group commutator decomposition of $\Delta$, i.e. $V$, $W$ such that $\Delta = V W V^\dagger W^\dagger$.

The Solovay-Kitaev algorithm is a recursive algorithm that approximates a given gate $U$ with an accuracy $\epsilon_n$. The algorithm exhibits convergence if $\epsilon_n < \epsilon_{n-1} < \ldots < \epsilon_0$. The idea of the proof can be explained as follows. Suppose we have a unitary $\Delta$, which is sufficiently close to identity matrix and is a balanced group commutator decomposition $\Delta = V W V^\dagger W^\dagger$, where the matrices $V$ and $W$ are also sufficiently close to the identity matrix. Then using the zeroth approximation for the matrices $V$ and $W$, we obtain the following approximation $\Delta_1 = V_0 W_0 V_0^\dagger W_0^\dagger$, which turns out to be better than if we used the zero approximation $\Delta_0$. This may seem counterintuitive, since the imperfectness of the zeroth approximation does not accumulate, but rather decreases.
Then we put $\Delta$ to be equal to the product $U U_0^\dagger$, where $U_0$ is the zero approximation for the matrix $U$. This matrix is close to unity. Then the first approximation $U_1 = V_0 W_0 V_0^\dagger W_0^\dagger U_0$ turns out to be better than the zeroth approximation. Subsequent recursion steps improve this approximation in a similar way. We refer the reader to the work \cite{dawson} for a full explanation and rigorous proof.

\section{$\mathcal{R}$-matrices as quantum gates}
\label{fund}

Important property of two-bridge $\mathcal{R}$-matrices (as a set of two diagonal matrices $T$ and $\bar{T}$ and two Racah matrices $S$ and $\bar{S}$ in comparison with most sets of universal quantum gates is that they can't be combined in any order. This leads to certain difficulties in calculating the sequences of $\mathcal{R}$-matrices corresponding to the different unitary matrices.

Another important property is that we do not have in this approach matrices, corresponding to one-qubit and entangling multi-qubit operations. Instead we use $\mathcal{R}$-matrices in higher representations as \textbf{qudit} operations. For example, in symmetric representation with the Young diagram $[3]$ all the matrices are of the size $4\times 4$ which means that they are qutetrit operations. But they, of course, can also be used to provide calculations in the case of a system of two qubits. In this paper we will provide an example of using quantum $\mathcal{R}$ matrices as quantum gates for fundamental representation, corresponding to one qubit operations.

Fundamental representation is the simplest case. In this case all the matrices are of the size $2\times 2$ \cite{symS,symS2}:

\begin{equation}
\begin{array}{l}
T=\left(\begin{array}{cc}q/A & \\ & -1/qA \end{array}\right),\ \ \ \bar{T}=\left(\begin{array}{cc}1 & \\ & -A \end{array}\right),
\\ \\
S=\frac{1}{\sqrt{(q+q^{-1})(A-A^{-1})}}\left(\begin{array}{cc}\sqrt{\frac{A}{q}-\frac{q}{A}} & \sqrt{Aq-\frac{1}{Aq}} \\ \sqrt{Aq-\frac{1}{Aq}} & -\sqrt{\frac{A}{q}-\frac{q}{A}} \end{array}\right),
\\ \\
\bar{S}=\left(\begin{array}{cc}\frac{q-q^{-1}}{A-A^{-1}} & \frac{\sqrt{(Aq-\frac{1}{Aq})(\frac{A}{q}-\frac{q}{A})}}{A-A^{-1}} \\ \frac{\sqrt{(Aq-\frac{1}{Aq})(\frac{A}{q}-\frac{q}{A})}}{A-A^{-1}} & -\frac{q-q^{-1}}{A-A^{-1}} \end{array}\right).
\end{array}
\end{equation}

These matrices are unitary if $A$ and $q$ are roots of unity. In the next section we will show how to use them as universal quantum gates. The quantum dimension of the fundamental representation shich is the final ingredient we need to find the knot polynomial is
\begin{equation}
D_{[1]}=\frac{A-A^{-1}}{q-q^{-1}}
\end{equation}

\section{Quantum gates through $\mathcal{R}$-matrices}

It is natural to expect that $\mathcal R$-matrices ~- constitute a set of universal gates, as they are elementary building blocks which constitute all the knots. In this section, we first present an efficient algorithm for finding the zeroth approximation of quantum gates by $\mathcal R$-matrices. Then, using this base, we use the Solovay-Kitaev algorithm to approximate some quantum gates that are widely used in quantum programming.

Of course, we can apply the Solovay-Kitaev algorithm in a straightforward manner. To do this, we should construct all sequences of $\mathcal{R}$-matrices up to length of $ l_0 $, striking from some representation (for example, $[1] \otimes [\bar1] \otimes [\bar1] \otimes [1] $) to the same representation. Going through these sequences, we should find the best zero approximation for the unitaries. After that we can apply the Solovay-Kitaev algorithm as usual. However, the set of sequences of zero approximations, so constructed, turns out to be extremely reducible, which leads to an increase in the compilation time of the quantum algorithm. One of the goals of the present work is that the irreducible basis of sequences of the zeroth approximation can be constructed and parameterized by knots.

Indeed, the 11-component product of two-bridge $\mathcal R$-matrices is nothing but the HOMFLY-PT polynomial $H^{\mathcal{K}} _ {[1]} (q, A) / D_{[1]}$ of the knot $\mathcal{K}$ (up to a phase $q^n$) obtained by a appropriate contraction of the braid corresponding to such a product. Since the two-bridge $\mathcal{R}$-matrices in the fundamental representation are $2\times 2$ unitary matrices, they have 3 independent parameters. Thus, one of the parameters of the product is fixed and is in one-to-one correspondence with a knot. This is the reason for the extreme reducibility of the set of sequences described above, since the same knot has a huge number of realizations through the products of $\mathcal R$-matrices. The remaining two parameters (phases) can be adjusted by multiplying the sequence corresponding to a fixed knot by additional $\mathcal R$-matrices and their inverses on the right and on the left without application of non diagonal Racah matrices. These matrices are diagonal, and do not change the knot invariant.

Thus, we provide the following implementation of the zeroth approximation function. First, one should construct an array of HOMFLY-PT polynomials of knots $\mathcal{K}$ up to some fixed number of self-intersections (these are known for a large number of knots and listed in the databases of knots \cite{katlas, knotebook}). After this, one should construct the sequences $S_{\mathcal{K}}$ of $\mathcal R$-matrices, which are representatives of each such knot. Now, in order to calculate the zeroth approximation for some gate $U$, we look for a knot from the array for which the difference between the absolute value of $U_{11}$ and the knot polynomial is minimal: $\mathcal{K} = \argmin_{\mathcal{K}'} | \, | U_ {11} | - | H^{\mathcal{K}'}_{[1]} (q, A) / D_{[1]} | \, |$. Then, one should multiply the sequence $S_{\mathcal{K}}$ corresponding to the found knot on the right and on the left by the $\mathcal{R}$-matrices so that the product will provide the best possible approximation of $U$.

The arguments in favor of the accuracy of the approximation of this algorithm are as follows. Consider the well-known parameterization of of a special unitary matrix
\begin{equation}
    U =
    \left(\begin{array}{cc}
    \cos \theta \, e^{\psi + \phi} & \sin \theta \, e^{\psi - \phi} \\
    \sin \theta \, e^{\phi - \psi} & \cos \theta \, e^{-\psi - \phi}
    \end{array}\right)
\end{equation}
One can observe that the absolute values of HOMFLY-PT polynomials divided by the quantum dimension $D_{[1]}$ rather densely fill the segment $[0, 1]$. This fact allows us to approximate the angle $\theta$ with good accuracy. Further, multiplying $S_{\mathcal K}$ by the powers of $\mathcal R$-matrices from the right and left allows one to trim the phases $\phi$ and $\psi$. The matrices are diagonal and nonzero elements are some degrees of $q = \exp \frac{2i\pi} {N + k} $, that is, the phase $\frac{2\pi}{N + k}$ sets the minimal approximation accuracy of $ \phi $ and $ \psi $. For relatively large $N + k$, the accuracy is sufficient for the zeroth approximation (note that $N + k \ge 2(N + 1) $ from the requirement of unitarity of $\mathcal R$-matrices).

Thus, we conclude that the natural irreducible basis of sequences of $\mathcal R$-matrices, which can be used for the zeroth order approximation of necessary quantum gate, can be parameterized by knots up to some fixed number of self-intersections. It also allows to compare the time of execution of the algorithm with the time of an algorithm that uses all sequences up to the fixed length $l_0$. The latter behaves like $O(3^{l_0}) $, which is much more than the number of knots with the number of self-intersections less than $l_0$ (e.g. 250 for $l_0 \le 10$). In addition, HOMFLY-PT polynomials are defined for arbitrary $q$ (and are known for knots up to 12 self-intersections \cite{katlas, knotebook}), which makes the presented algorithm universal for any dimensions of group $N$ and level $k$.

Finally, let us provide the results of applying the algorithm to the Hadamard gate $H$ and the so-called $\pi/8$-gate $E$
\begin{equation}
    H = \frac1{\sqrt{2}} \left(
    \begin{array}{cr}
        1 & 1 \\ 1 & -1
    \end{array}\right), \quad
    E =  \left(
    \begin{array}{cc}
        1 & 0 \\ 0 & e^{\frac{i\pi}4}
    \end{array}\right)
\end{equation}

\begin{figure}[h!]
\centering
\begin{center}
\includegraphics[width=0.45\textwidth]{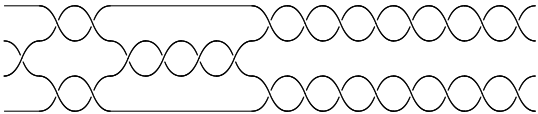}
\end{center}
\caption{1. Zeroth approximation for Hadamard gate, knot $9_4$, accuracy $\epsilon = 0.043,\ N = 2,\ k = 13$}
\end{figure}

\begin{figure}[h!]
\centering
\begin{center}
\includegraphics[width=0.45\textwidth]{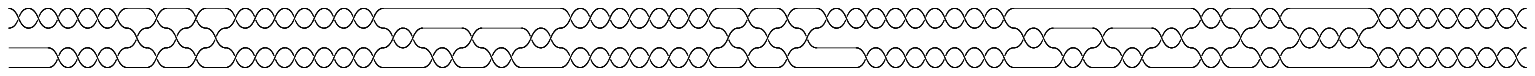}
\end{center}
\caption{2. First approximation for Hadamard gate, knots $7_4$, $9_{23}$, $7_4$, $9_{23}$, $9_4$, accuracy $\epsilon = 0.0068,\ N = 2,\ k = 13$}
\end{figure}

\begin{figure}[h!]
\centering
\begin{center}
\includegraphics[width=0.45\textwidth]{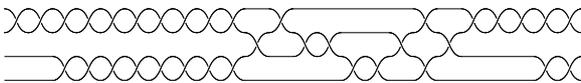}
\end{center}
\caption{3. Zeroth approximation for $\pi/8$-gate, knot $10_{29}$, accuracy $\epsilon = 0.028,\ N = 2,\ k = 9$}
\end{figure}

\begin{figure}[h!]
\centering
\begin{center}
\includegraphics[width=0.45\textwidth]{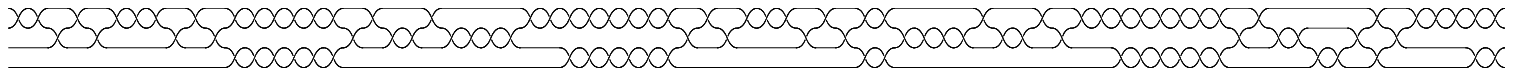}
\end{center}
\caption{4. First approximation for $\pi/8$-gate, knots $9_{17}$, $9_{11}$, $9_{17}$, $9_{11}$, $10_{29}$, accuracy $\epsilon = 0.005,\ N = 2,\ k = 9$}
\end{figure}
Examples of the approximations of the indicated gates in the zeroth and first order are displayed on figures 1-4. The structure of the zeroth approximation has a clear interpretation in terms of the proposed algorithm: the braid corresponding to the particular implementation of the knot is ``dressed'' in $\mathcal R$-matrices on the left and on the right. The structure of the first approximation is consistent with Solovay-Kitaev algorithm and has the form $U_1 = V_0 W_0 V_0^\dagger W_0^\dagger U_0$.

Unfortunately, the presented algorithm is not directly applicable to the higher symmetric representations $[r]$. The reason is that now the knot polynomial fixes only one of $(r+1)^2 - 1$ components of the unitary matrix. This brings us back to the requirement of using zeroth approximation method that involves all the sequences up to some fixed length $l_0$. In addition, according to estimates of \cite{dawson}, the length of the sequence $l_0$ should be significantly larger than in the case of $r = 1$. This forces us to resort to high-performance computing methods. Thus the studied methods require further modifications to apply them to larger matrices. This is currently a work in progress.

\section{Conclusion}

In this letter we used quantum $\mathcal{R}$-matrices as universal quantum gates. This provides an approach to the topological quantum computer which is natural from the point of view of Chern-Simons theory and knot theory. This means that each quantum algorithm and program correspond to some knot. We use the Solovay-Kitaev algorithm to provide such an approximation to one-qubit operations. We do it using the examples of Hadamard and $\pi/8$-gate. These approximations use the $\mathcal{R}$-matrices and Racah matrices for the fundamental representation.

This approach can be generalized to higher representations and thus larger unitary matrices and qudit or multiqubit operations. However there are certain technical difficulties in applying the algorithm to larger matrices, which at the moment are unsolved. Another interesting question which remains to be studied is how to move from two-bridge knots to other types of knots.

\section*{Acknowledgements}

We are grateful for very useful discussions to A. Andreev, N. Tselousov, S. Mironov, A. Sleptsov and A. Popolitov.

This work was supported by Russian Science Foundation grant No 18-71-10073.

\end{document}